# Stability Engineering of Halide Perovskite via Machine Learning

Zhenzhu Li[1,2], Qichen Xu[1,2], Qingde Sun[1,2], Zhufeng Hou[3], Wan-Jian Yin[*,1,2]

[1]Soochow Institute for Energy and Materials InnovationS (SIEMIS), College of Physics, Optoelectronics and Energy & Collaborative Innovation Center of Suzhou Nano Science and Technology, Soochow University, Suzhou 215006, China

[2]Key Laboratory of Advanced Carbon Materials and Wearable Energy Technologies of Jiangsu Province, Soochow University, Suzhou 215006, China

[3]Center for Materials research by Information Integration, National Institute for Materials Science, 1-2-1 Sengen, Tsukuba, Ibaraki 305-0047, Japan

Perovskite stability is of the core importance and difficulty in current research and application of perovskite solar cells. Nevertheless, over the past century, the formability and stability of perovskite still relied on simplified factor based on human knowledge, such as the commonly used tolerance factor *t*. Combining machine learning (ML) with first-principles density functional calculations, we proposed a strategy to firstly calculate the decomposition energies, considered to be closely related to thermodynamic stability, of 354 kinds halide perovskites, establish the machine learning relationship between decomposition energy and compositional ionic radius and investigate the stabilities of 14,190 halide double perovskites. The ML-predicted results enable us to rediscover a series of stable rare earth metal halide perovskites (up to ~$10^3$ kinds), indicating the generalization of this model and further provide elemental and concentration suggestion for improving the stability of mixed perovskite.

The emergence of new applicable materials can often promote the long-lasting development of a particular field in modern science and technology. As a typical example, recent years have witnessed a surge of research interest in solar cell field[1-4], which is rooted to the discovery of cutting-edge halide perovskite materials. Perovskites solar cells (22.7%) have become a new front runner in the race of cell efficiency and surpassed CdTe (22.1%) and $Cu_2(In,Ga)Se_2$ (22.6%)[5], major contenders in the thin-film solar cell industry. The main remaining obstacle for the commercialization of perovskite solar cell is the long-term stability. Although great efforts are made in cell architecture and encapsulation, the ultimate way is to improve the intrinsic stability of perovskite materials. Apart from photovoltaics, stable halide perovskites also exhibit their great potential in opt-electric applications, such as wide-spectrum light-emitting-diode[6] and high-sensitivity X-ray detector[7].

Prototype $CH_3NH_3PbI_3$ is easily decomposed into secondary phases such as $CH_3NH_3I$, $PbI_2$ and $I_2$ under moisture, air, temperature and light, leading to fast degradation of cell performance, which is reflected by its almost neutral decomposition energy[8]. To address the stability issue, compositional management of perovskite $ABX_3$ leads to tremendous mixed perovskites with double or multiple elemental mixing[9-19] of (K, Rb, Cs, MA, FA), (Pb, Sn, Cd, Mn) and (I, Br, Cl) on A, B and X sites respectively, demonstrating enhanced stability in comparison to their single perovskite counterpart. For example, although the photo-active perovskite phase of $RbPbI_3$, $CsPbI_3$ and $FAPbI_3$ are unstable at room temperature, proper mixing of (Rb, Cs, FA) on A-site can result in stable mixed-A-site perovskite[15,17]. Now, precise control of stability in experiment aims to utilize more mixing elements such as triple-A $(Cs,MA,FA)InBiBr_6$[20], triple-A double-X $Cs_x(MA_{0.17}FA_{0.83})_{(1-x)}Pb(I_{0.83}Br_{0.17})_3$[14] and quadruple-A double-X Rb-$FA_{0.75}MA_{0.15}Cs_{0.1}PbI_2Br$[21], making the problem more complicated. Fundamental insight is urged to understand the effect of elemental mixing and provide guidance for stability engineering *i.e.* the type of mixing elements and their concentrations.

Apart from mixed perovskite, the class of double halide perovskites with a formula

$A_2B(I)B(III)X_6$ that can be considered as the split of B-cation in $AB(II)X_3$ to monovalent B(I) and trivalent B(III) cations, offer vast perovskites candidates up to ~$10^4$, opening up a new treasure trove of materials for resolving current stability issue[22]. $Cs_2AgBiX_6$ [X=Br, Cl] is the first double halide perovskites that was successfully synthesized via a solid state or solution route for potential photovoltaic applications in early 2016[23,24], stimulating the experimental synthesis of more double halide perovskite such as $Cs_2AgInCl_6$, $Cs_2NaB(III)Cl_6$ [B(III) = In, Sb, Bi, Tl, Fe, Ti], $Cs_2KB(III)Cl_6$ [B(III) = In, Bi] [see Table S1]. Meanwhile, first-principles calculations based on density functional theory (DFT) have become a powerful tool to screen stable halide perovskites for photovoltaic application[25,26]. So far, experimental and theoretical efforts have investigated only up to ~$10^2$ kinds of materials, far less than the possible candidates amount of ~$10^4$. Meanwhile, none of the explored ~$10^2$ perovskites has promise to be a stable and high-efficiency solar cell absorber. Therefore, a high-throughput fast-screening methodology is required to speed up the current discovery of new stable perovskites.

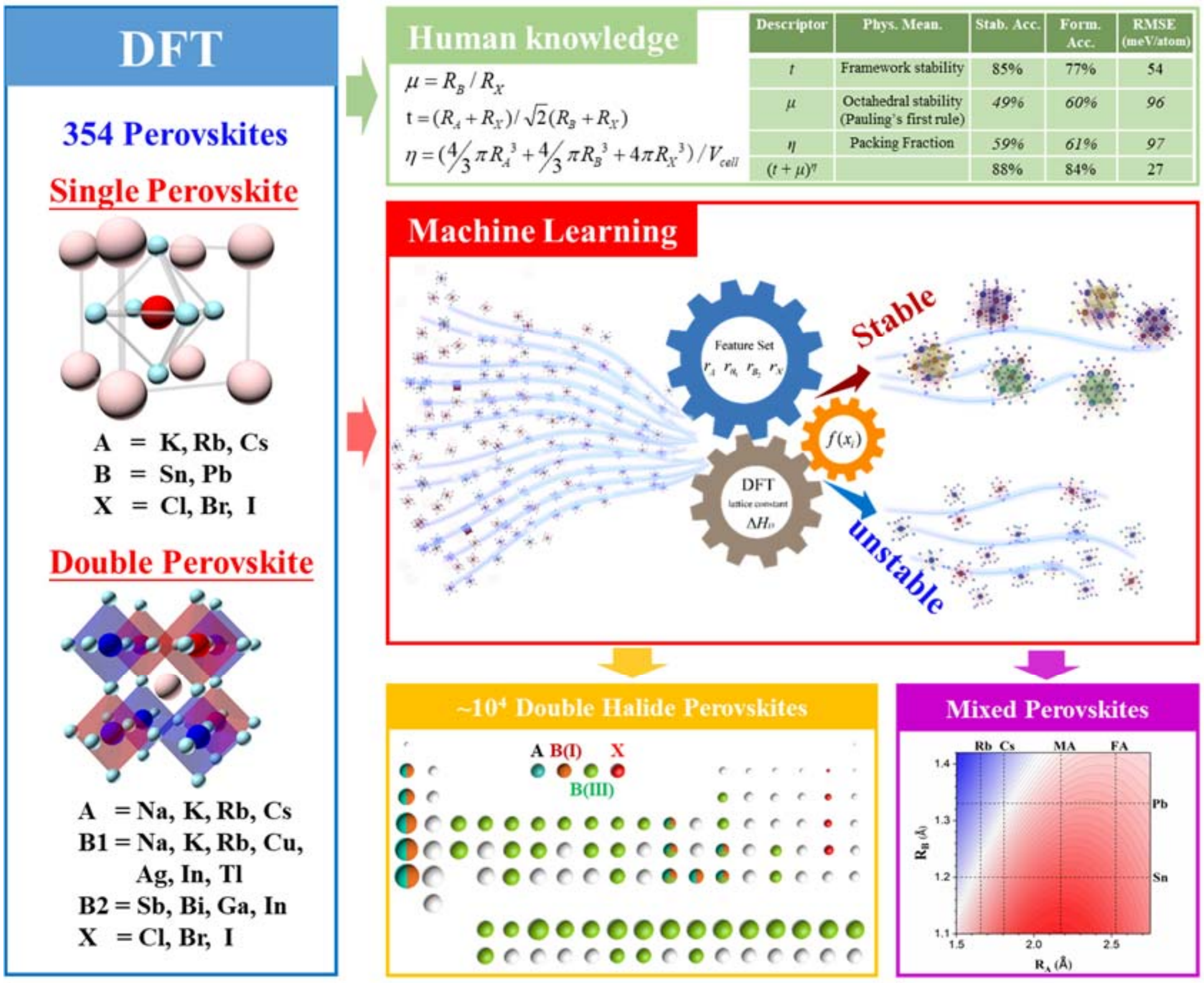

**Figure 1. Schematic strategy of ML based on DFT for high-throughput stability engineering of halide perovskites.** The decomposition energies ($\Delta H_D$) of 354 perovskites are calculated by DFT and then used as data sets for machine-learning model, in comparison to the performance of conventional human-knowledge descriptors, to predict the stability of 14190 double halide perovskites and provide the elemental and compositional landscape for stability of mixed perovskites.

### Perovskite stability: human knowledge .vs. machine learning

Although the perovskite stability is a long-standing topic, the practical problems above have made one's desire for stable perovskites never as urgent as now. Conventional materials discovery has been driven by a trial-and-error process due to the limited capability of experiments and theoretical tools. Owing to the simple crystal structure of perovskites, tolerance factor $t$, based on human cognition of cubic geometry[27] [Figure 1], has become a popular stability descriptor and accelerated the qualitative screening of stable perovskites during the last 100 years. Nevertheless, its quantitative accuracy for the formability and stability is actually not so good as an efficient descriptor on precisely engineering stable halide perovskite[28,29]. Considering $t$ only describe the stability of general structural framework *i.e.* cubic $ABX_3$ phase with corner-sharing octahedron [Figure 1], we recently proposed a new stability descriptor $(\mu+t)^\eta$ by further considering the stability of $BX_6$ octahedron (octahedral factor $\mu$) and atomic packing fractions ($\eta$)[29]. The accuracy of relative stability for descriptor $(\mu+t)^\eta$ is up to 90%, in comparison to ~70% for $t$ within the model systems of 138 perovskites. In fact, all $t$, $u$, $\eta$ and $(\mu+t)^\eta$ are based on functions of compositional ionic radii, considered as key features correlated with perovskite stability. In mathematical words, one is trying to construct a mapping between perovskite stability (*property*) and compositional ionic radius (*feature*) [Figure 1] with the mapping formula derived from human knowledge of physics and chemistry on perovskite stability *i.e.* structural framework ($t$), Pauling's first rule ($\mu$) and atomic packing fraction ($\eta$). Recent researches and applications have proven that such kind of *feature-properties* mapping can be done more efficiently and precisely by artificial intelligence or machine learning, which demonstrated remarkable learning, prediction and decision ability in diverse areas such as materials design[30-32], chemical reaction[33], medical diagnosis[34] as well as Go game[35].

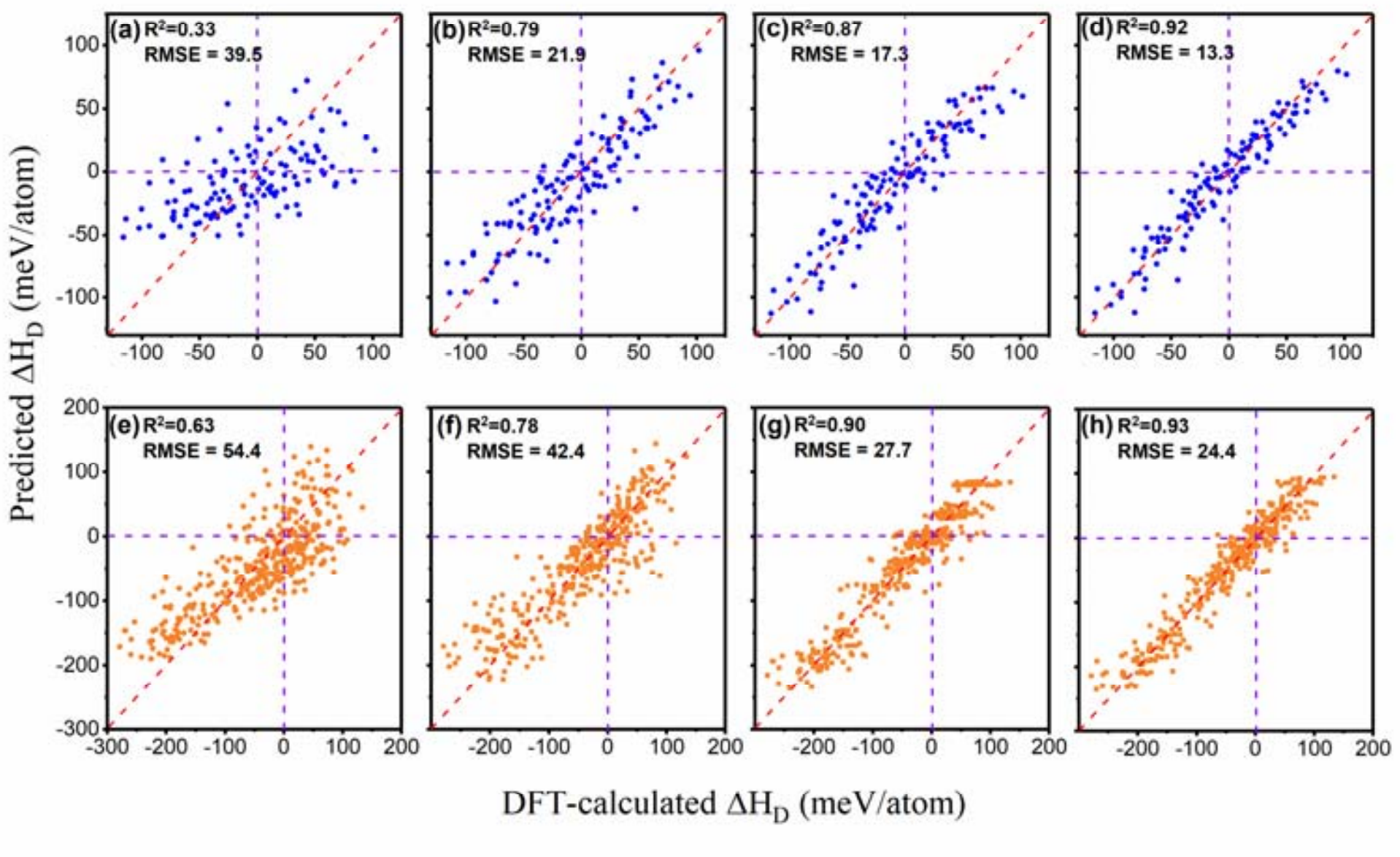

**Figure 2. Fitting performance of human knowledge .vs. machine-learning.** DFT-calculated $\Delta H_D$ .vs. (a)/(e) descriptor-$t$-predicted, (b)/(f) descriptor-$(\mu+t)^\eta$-predicted $\Delta H_D$, and machine learning predicted $\Delta H_D$ based on the features of (c)/(g) ($R_A$, $R_{B,eff}$ and $R_X$) and (d)/(h) ($R_A$, $R_{B1}$, $R_{B2}$, $R_X$) for 126/354 perovskites investigated in previous[29]/current work.

Therefore, we build a strategy to combine first-principles DFT calculations with machine-learning for high-throughput screening of stable perovskites among ~$10^4$ perovskites. First, we calculated by DFT, the decomposition energies of 354 halide perovskites, including single and double ones as composition shown in Figure 1. After that, proper machine learning model is trained based on the feature input ($R_A$, $R_{B1}$, $R_{B2}$, $R_X$) and properties output ($\Delta H_D$) of 354 sets of data. This model is then used to predict the $\Delta H_D$ for other ~$10^4$ halide perovskites and mixed-halide perovskites, which then provides insights and guidance on experiments.

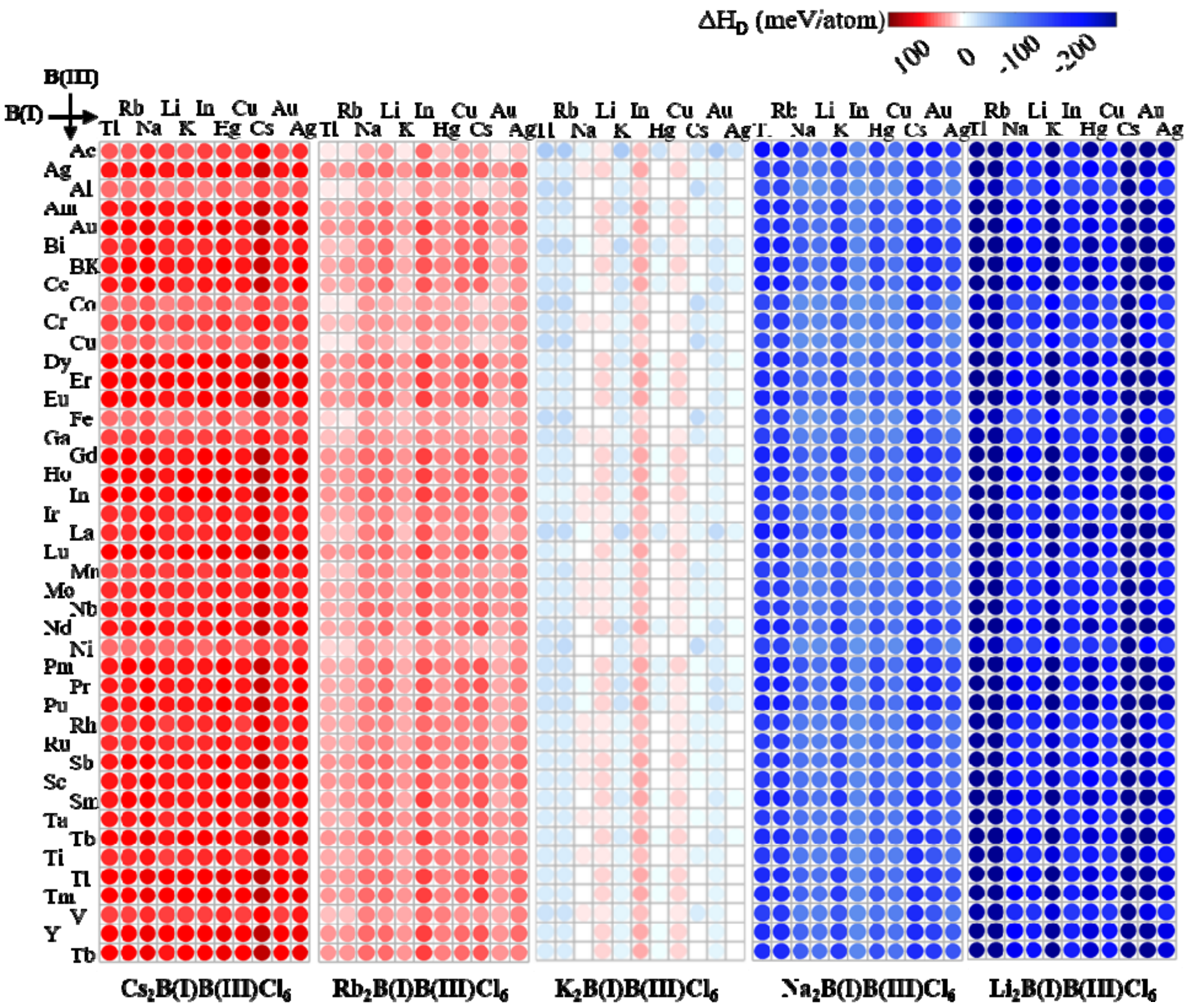

**Figure 3. Heat map of decomposition energies of 2365 perovskites with A=Cs, Rb, K, Na, Li and X=Cl.** The red/blue color indicates the positive/negative decomposition energies and the absolute value is presented by color shades.

To demonstrate the prediction power of machine learning, we firstly choose our previous data[29] on decomposition energies ($\Delta H_D$) of 126 kinds of double halide perovskites. Using $t$ and $(\mu+t)^\eta$ as descriptor, the root-mean-square-error (RMSE) of predicted $\Delta H_D$ are 39 meV and 22 meV and their prediction accuracies for formability (relative stability) are 68% (80%) and 85% (90%) [Figure 1]. In double halide perovskite $A_2B(I)B(III)X_6$, $t$, $u$ and $\eta$ are functions of ionic radius $R_A$, $R_{B,eff}$ and $R_X$, where $R_{B,eff} = (R_{B(I)}+R_{B(III)})/2$. To be fair, we choose the same feature set ($R_A$, $R_{B,eff}$ and $R_X$) for machine learning, which produce a mapping between property ($\Delta H_{D,ML}$) and features ($R_A$, $R_{B,eff}$ and $R_{,X}$). The RMSE of $\Delta H_{D,ML}$ to $\Delta H_{D,DFT}$ is 17 meV and the prediction accuracies for formability (relative stability) is 89% (93%), indicating machine learning can establish a better mapping between radius and stability than the descriptor $(\mu+t)^\eta$ that we proposed in previous work[29]. In the present work, we expanded our DFT calculations to 354 kinds of halide perovskites and the results verify the conclusions that machine-learning works better than $(\mu+t)^\eta$ and $(\mu+t)^\eta$ works better than $t$.

For double halide perovskites $A_2B(I)B(III)X_6$, conventional descriptors, whether $(u + t)^\eta$ or $t$,

treat B(I) and B(III) as the same and the effective radius ($R_{eff}$) is ($R_{B1}+R_{B2}$)/2. Therefore, they cannot distinguish two perovskites with the same ($R_{B1}+R_{B2}$)/2. For example, machine learning with feature on ($R_A$, $R_{B,eff}$ and $R_X$) lead to almost the same decomposition energies for $Cs_2AgBiCl_6$ (81.2 meV/atom), $Cs_2KInCl_6$ (81.7 meV/atom) and $Cs_2RbGaCl_6$ (79.9 meV/atom), since the sum of radius of $Ag^{1+}$(1.29 Å) and $Bi^{3+}$(1.17 Å), $K^{1+}$(1.52 Å) and $In^{3+}$(0.94 Å), $Rb^{1+}$(1.66 Å) and $Ga^{3+}$(0.76) are almost the same. The situations are much improved when $R_{B(I)}$ and $R_{B(III)}$ are treated as two different features for machine learning model [Figure 2(h)]. Statistically, when considering $R_{B(I)}$ and $R_{B(III)}$ separately, all the RMSE, prediction accuracy of relative stability and formability have improved. Therefore, we choose ($R_A$, $R_{B1}$, $R_{B2}$, $R_X$) as features in our machine learning models for predicting stable perovskites in the following.

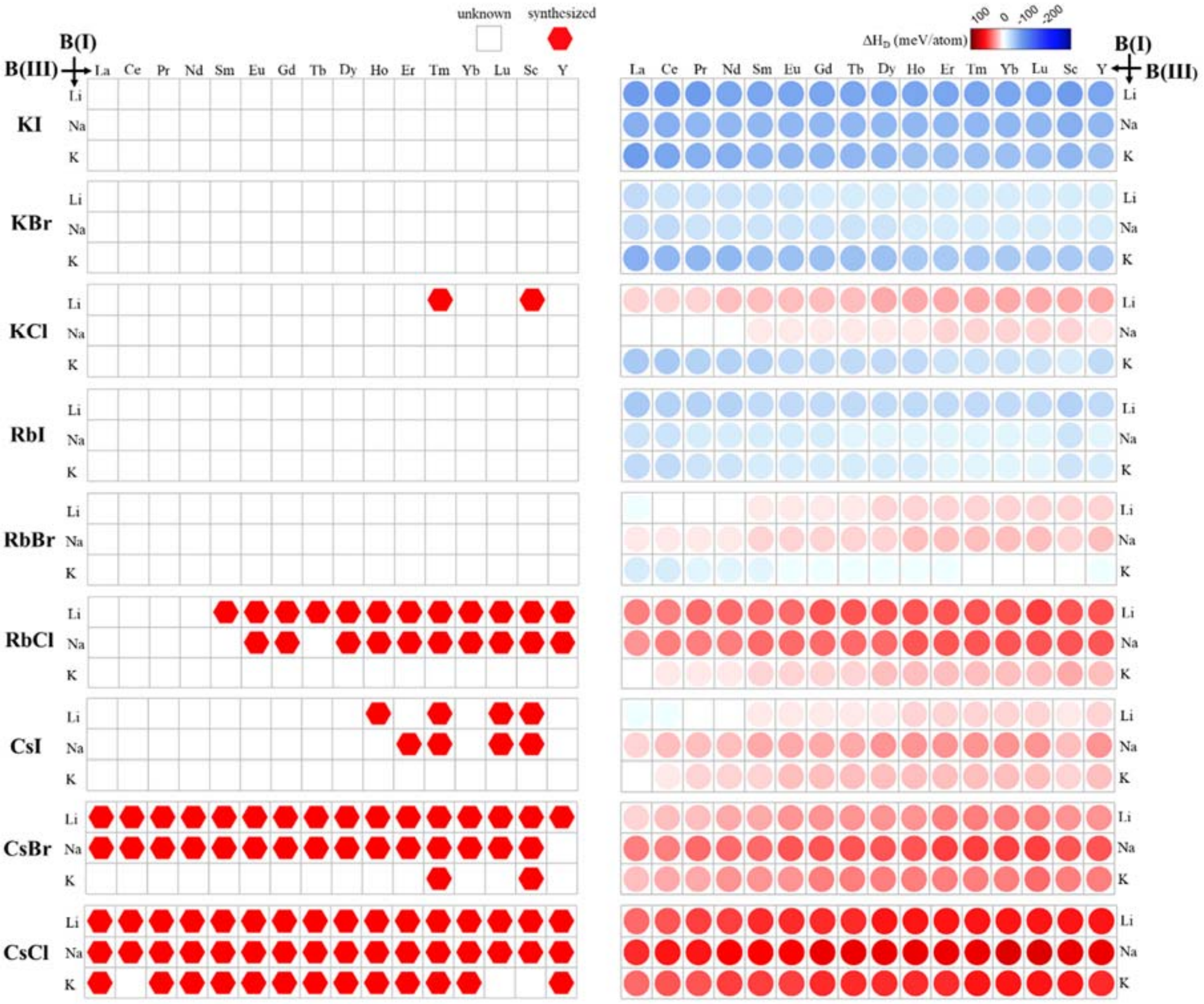

**Figure 4. The comparison of experiments and ML results for rare-earth-element-related double halide perovskites.** Left panels are experimental results based on the reference [36]. Right panels are current ML results. A and X elements are listed on the left.

### Exploration of stable double halide perovskite by machine learning

The machine learning model is then used to predict the decomposition energies of 14,190 kinds of double halide perovskites $A_2B(I)B(III)X_6$ with 10 kinds of A elements, 11 kinds of B(I) elements, 43 kinds of B(III) elements and 3 kinds of X elements, as compositions shown in the inserted periodic table of Figure 1 and results shown in Figure 3 and S1-S5. Here, we restrained our searching space within double halide perovskites $A_2B(I)B(III)X_6$, where A, B(I) and B(III) are 1+, 1+ and 3+

cations and X is 1- halide ions, since most of perovskites in training set are in the same class. Among 14,190 perovskites, 2,275/11,915 kinds are predicted to be stable/unstable ($\Delta H_D$ >/< 0). As our calculated/predicted decomposition energy is the upper limit value [see Methods part], the number of stable perovskite should be even less. The most stable AX class [$A_2B(I)B(III)X_6$] is CsCl with 473 kinds of perovskites predicted to stable with average $\Delta H_D$ of 82.7 meV/atom, following by RbCl (473 kinds, 44.6 meV/atom), CsBr (448 kinds, 31.5 meV/atom), KCl (247 kinds, 0.1 meV/atom), AuCl (217 kinds, -3.4 meV/atom), RbBr (219 kinds, -4.1 meV/atom) and CsI (198 kinds, -10.8 meV/atom). Other perovskites including all with Li, Na, Cu, Ag, Hg, In as A-site are unstable. The heat map of decomposition energy of CsCl-, RbCl-, KCl-, NaCl-, LiCl-class perovskites are shown in Figure 3. Clear stability trend is observed from CsCl- to LiCl-class, where KCl is the critical region where $\Delta H_D$ is around zero. Similar neutral regions for bromide and iodide are RbBr and CsI respectively, which explains the difficulty in searching stable iodide double perovskites for low-bandgap sunlight absorber[23,37].

The results of recent synthetic attempt for double halide perovskites are listed in Table S1, which shows that DFT and ML are correct to predict the formability of most perovskites. For those without DFT results, ML can also predict the formability of most perovskites, such as abnormal $Cs_2Au^{1+}Au^{3+}I_6$, which was recently proposed to be efficient single-junction solar cell absorbers[38]. The major discrepancies come from perovskites with $Cu^{1+}$ and $In^{1+}$ at B(I) site. This is because for $Cu^{1+}$, the covalent nature of its high-level 3*d* orbital make it favor four-coordinate instead of six-coordinate in perovskites[39] and for $In^{1+}$, it is extremely easily oxidized to its stable charge state $In^{3+}$ [40]. To completely consider the stability of those perovskites, additional pathways including competing phases with more coordinates and charge states should be considered, which is out of the scope of current work. For those perovskites, our ML results are in agreement with the DFT results both from ours and others, indicating that this discrepancy is not derived from ML methods but from the reliability of DFT data compared to experiments.

One of the major goals for machine learning in materials design is to find new materials out of training sets therefore speed up materials discovery and guide the experiment. One of the intriguing results here is that current ML model successfully predict the stability trend of thousands of rare-earth-element-related double halide perovskite $A_2B(I)B(III)X_6$ with B(I) = Li, Na, K and B(III) = Sc, Y, La, Ce, Pr, Nd, Sm, Eu, Gd, Tb, Dy, Ho, Er, Tm, Yb, Lu. In Figure 4, we can see that (*i*) Machine learning predicted the clear trend of experimental formability from KI- to CsCl-class perovskites; (*ii*) For individual perovskites, machine learning successfully predicted the stable perovskites that has been experimentally synthesized and none of the perovskites predicted to be unstable has been synthesized. It is noted that recent interest in double halide perovskite initiated in 2016 when the $Cs_2AgBiBr_6$ was synthesized[23,24]. The class of rare-earth-element-related double halide perovskite has drawn rarely attention currently and therefore been unexplored yet. The ML results have driven us to rediscover those large quantities of stable double halide perovskites that have been successfully synthesized in 1960*s*-1980*s*[36]. The consistency as shown in Figure 4 partially prove the generalization of our ML model, which may provide guidance for experiments to explore much more stable perovskites that have not been synthesized as shown in Figure 3 and S1-S5.

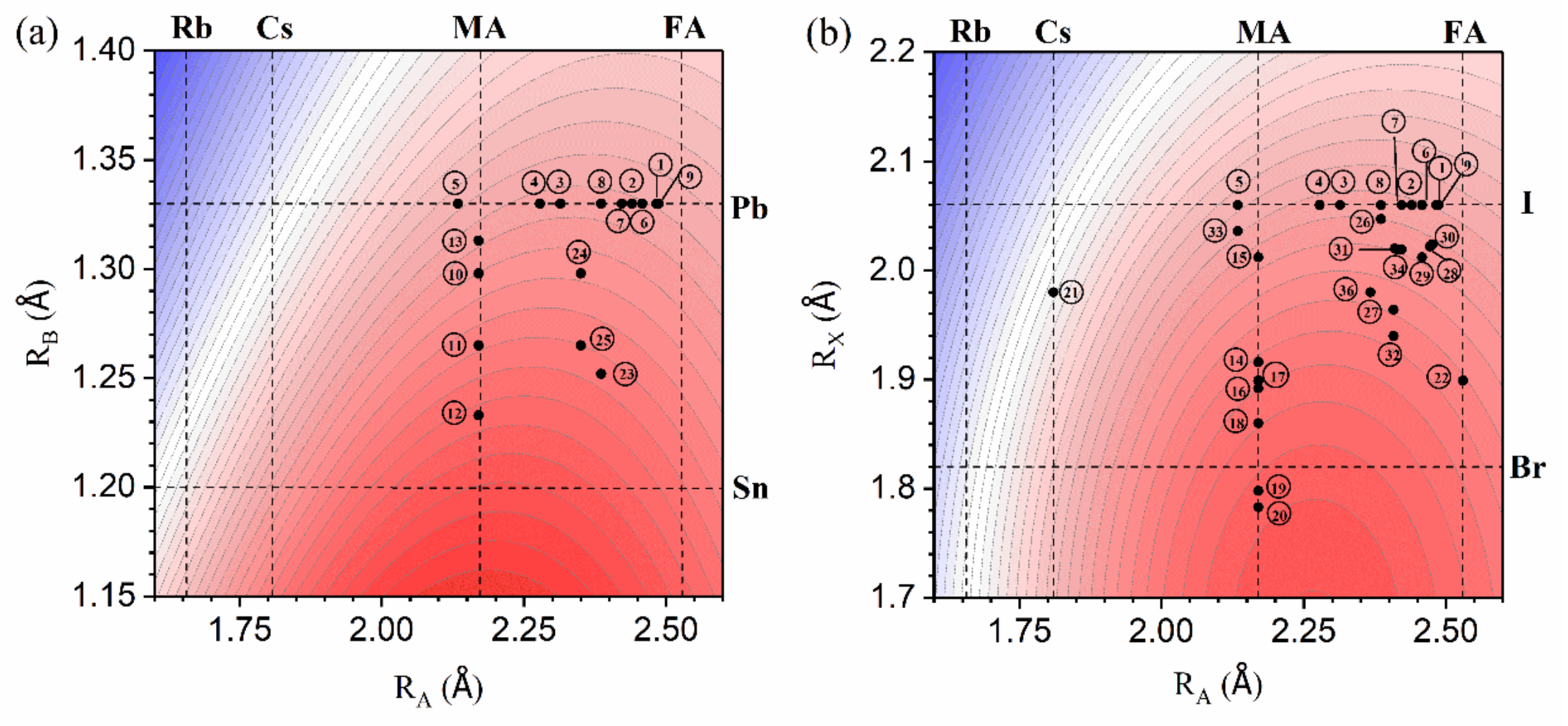

**Figure 5. The ML-predicted decomposition energies for mixed perovskites (a) $ABI_3$ and (b) $APbX_3$.** The x/y axis are ionic radii of A/B and A/X in (a) and (b), respectively. The available experimental results are marked with their compositions and references in Table S2.

**Compositional engineering to stabilize perovskite**

One advantage of machine-learning methods is its convenience to explore the landscape of stability on continuous feature space therefore capable of providing valuable information and guidance for the stability of mixed-elements perovskites. In contrast, first-principles DFT calculations or individual experiments provide only scattered points as shown in Figure 5 of *feature-properties* plot and therefore difficult to investigate the full landscape on feature space.

It is observed in Figure 5 that (*i*) the perovskite stability can be enhanced by chemical mixing of different cations and anions on all the A, B and X site. For example, the enhanced stability of $MA_xFA_{1-x}PbI_3$ is reflected by the convex curve in the whole range of $0 < x < 1$ of both Figure 5(a) and (b). Also, the addition of Br in $APbI_3$ (A = Cs, MA, FA) resulting in $APb(I_{1-x}Br_x)_3$ can enhance the perovskite stability; (*ii*) Although $Sn^{2+}$ is easily oxidized, the addition of Sn at B-site of $(MA/FA)PbI_3$ can make $(MA/FA)(Pb_{1-x}Sn_x)I_3$ have better stability than its pure Pb-counterpart $(MA/FA)PbI_3$, which could be a viable way to engineer stable low-bandgap perovskites for higher efficiency single junction solar cell[41,42] and perovskite tandem cell[43]; (*iii*) Many of existing experiments are focused on the chemical mixing of double or multiple elements on one particular site. Taking A-site mixing as an example, much efforts are made on $(MA,FA)PbI_3$, $(Cs,MA)PbI_3$, $(Cs,FA)PbI_3$ and $(Rb,FA)PbI_3$[12-15,44]. In principles, those mixing has the same effect, *i.e.* tuning the effective radius of A site. For example, the positions of ①$MA_{0.13}FA_{0.87}PbI_3$ in Figure 5 is almost the same as ⑨$Rb_{0.05}FA_{0.95}PbI_3$. From this perspective, the effect of multiple mixing on single site can be achieved by double mixing; (*iv*) Most of current experiments focus on single-site double/multiple-mixing which is constraint in particular lines (dashed lines in Figure 5) therefore limits the tunability. Alternatively, wide unexplored region in Figure 5 with great potential for high stability can be arrived by double-elements multiple-sites mixing. It is noted that in the last five years, the stable perovskites absorbers for world-record cell efficiency has been developed[5,45] from pure phase $MAPbI_3$, to single-site double-mixing [$MAPb(I_{1-x}Br_x)_3$(x=0.1-0.15)], to double-site double mixing [$FA_{0.85}MA_{0.15}PbI_{2.55}Br_{0.45}$, $(FAPbI_3)_{0.95}(MAPbBr_3)_{0.05}$], to double-site triple-mixing [$Cs_{0.1}FA_{0.75}MA_{0.15}PbI_{2.49}Br_{0.51}$] then back to double-site double mixing [$(FAPbI_3)_{0.95}(MAPbBr_3)_{0.05}$].

The results in Figure 5 provide directions to engineer more stable mixed perovskites for which a balance for optical and electronic properties are required as high-efficiency stable perovskite solar cell absorbers.

## Conclusion

We proposed a strategy by combining machine-learning and first-principles DFT calculations to engineer the stability of halide perovskites. By choosing DFT results of 354 perovskites as training set, we established a machine learning mapping between perovskite stability and ionic radius, which demonstrated better performance than the descriptors of long-lasting tolerance factor *t* and recently proposed $(\mu+t)^{\eta}$. This machine learning model is then used to predict the stability of over $10^4$ halide double perovskites, among which up to ~$10^3$ rare-earth-metal perovskites are rediscovered. The stability trend predicted by machine learning model is in great agreement with available experiments, indicating the generalization capability of current machine model. Synthesis are called for the unexplored perovskites that are predicted to be stable in this model, therefore providing guidance for experiments. Machine-learning model is then used to predict the stability landscape of mixed perovskites, rationalize diverse experimental results in the field and suggest double-elements multiple-sites mixing strategies for further improving the stability of mixed perovskite.

## Methods

### Decomposition energy calculations

For double halide perovskite $A_2B(I)B(III)X_6$, the common decomposition pathway, $A_2B(I)B(III)X_6 \rightarrow 2AX+B(I)X+B(III)X_3$ was considered in first-principles DFT calculations. This pathway was chosen to study the thermodynamic stability of 126 halide perovskites and a stability trend was found, which did not have significant change when additional pathways are considered[29]. The large discrepancies come from certain perovskites such as $Cu^{1+}$ and $In^{1+}$-based ones that were discussed in text. When additional pathways are considered, the decomposition energies may decrease. Therefore, the decomposition energies predicted in this work are upper limit values.

The total energies of perovskites and their secondary phases were calculated using the computational package VASP[46]. The projector augmented-wave (PAW)[47] pseudopotential with the general gradient approximation (GGA) of the PBEsol exchange correlation functional was adopted, and the cutoff energy was 300 eV. The same k-mesh density was chosen for all the structures by setting KSPACING = 0.12. The geometry was relaxed until all atomic forces were less than 0.01 eV/Å, and at each self-consistent field iteration the convergence criterion for the total energy was 0.5 meV.

### Machine-learning method

The hidden relationship between the output decomposition energy ($\Delta H_D$) and the input feature space (***d***) is achieved in terms of our supervised machine learning process. In detail, Kernel Ridge Regression (KRR)[48] with a Radical Basis Function ('rbf') kernel implemented in scikit-learn 0.19.1 package[49] is employed for our machine learning model. Theoretically, for a Gaussian kernel $k(x,x_i)=\exp(-\frac{\|x-x_i\|^2}{2\sigma^2})$, $\Delta H_D$ is expressed as a weighted sum of Gaussians: $\Delta H_D=\sum_{i=1}^{N} c_i \exp\left(\frac{-\|d_i-d\|_2^2}{2\sigma^2}\right)$, where $N$ is the number of training data points. The coefficients $c_i$ are

determined by minimizing $\sum_{i=1}^{N}[\Delta H_{D,\text{predicted}} - \Delta H_{D,\text{calculated}}]^2 + \lambda \sum_{i,j=1}^{N,N} c_i c_j \exp(-\|d_i - d_j\|_2^2 / 2\sigma^2)$, where $\|d_i - d_j\|_2^2 = \sum_{\alpha=1}^{\Omega}(d_{i,\alpha} - d_{j,\alpha})^2$ is the squared $l_2$ norm of the difference of descriptors of different materials, i.e., their "similarity", $\lambda$ controls the strength of the regularization and is linked to the noise level of the learning problem, and $\sigma$ is the length scale of the Gaussian, which controls the degree of correlation between training points.. In our case, the best squared-$l_2$-norm regularization hyperparameters '$\lambda$' and '$\sigma$' are chosen separately inside a parameter-grid with the help of 5-fold lave-one-out cross-validation, *i.e.*, by leaving some of the calculated materials out in the training process and testing how the predicted values for them agree with the actually calculated ones, for example here, root-mean-square error (RMSE) is adopted to evaluate the predictive ability of an output model. After getting the best training model over our 354 sets of data, prediction is made by using this model to output the $\Delta H_D$ of halide perovskites and mixed-halide perovskites covering a wide range (~$10^4$ combinations) of the periodic table.

## Acknowledgement

W.Y. thank Prof. Xingao Gong for insightful discussion and acknowledge the funding support from National Key Research and Development Program of China under grant No. 2016YFB0700700, National Natural Science Foundation of China (under Grant No. 11674237, No. 51602211), Natural Science Foundation of Jiangsu Province of China (under Grant No. BK20160299) and Suzhou Key Laboratory for Advanced Carbon Materials and Wearable Energy Technologies, China. The work was carried out at National Supercomputer Center in Tianjin and the calculations were performed on TianHe-1(A).

## Contributions

W.Y. conceived the idea and supervised the project. Z.L., Q.X., Q.S. and W.Y. performed the calculations. All authors contribute to the data analysis, discussions and the writing of the paper.

## Additional Information

The supporting information includes the summary of available experimental results on double halide perovskites and our DFT-calculated and ML-predicted decomposition energies (Table S1) and the mixed perovskites (Table S2) marked in Figure 5. Figure S1-S5 provide heat maps of decomposition energies of other 11,825 double halide perovskites, together with 2365 kinds shown in Figure 3 to a total of 14,190 kinds.

## Data Availability

All the features and DFT calculated decomposition energies for 354 kinds of halide perovskites are provided online. The machine learning predicted decomposition energies of 14190 double halide perovskites are provided online.

## Corresponding author

Correspondence to: Wan-Jian Yin (wjyin@suda.edu.cn)

## References

1. Kojima, A., Teshima, K., Shirai, Y. & Miyasaka, T. Organometal halide perovskites as visible-light sensitizers for photovoltaic cells. *J. Am. Chem. Soc.* **131**, 6050-6051 (2009).
2. Kim, H. S. *et al.* Lead iodide perovskite sensitized all-solid-state submicron thin film mesoscopic solar cell with efficiency exceeding 9%. *Sci. Rep.* **2**, 591 (2012).

3. Lee, M. M., Teuscher, J., Miyasaka, T., Murakami, T. N. & Snaith, H. J. Efficient hybrid solar cells based on meso superstructured organometal halide perovskites. *Science* **338**, 643-647 (2012).
4. Burschka, J. *et al.* Sequential deposition as a route to high-performance perovskite-sensitized solar cells. *Nature* **499**, 316-319 (2013).
5. NREL, Best Research Cell-Efficiencies, http://www.nrel.gov/ncpv/images/efficiency_chart.jpg.
6. Song, J. *et al.* Quantum dot light-emitting diodes based on inorganic perovskite cesium lead halides ($CsPbX_3$). *Adv. Mater.* **27**, 7162-7167 (2015).
7. Pan, W. *et al.* $Cs_2AgBiBr_6$ single-crystal X-ray detectors with a low detection limit. *Nature Photon.* **11**, 726-732 (2017).
8. Zhang, Y. Y., Chen S. Y., Xu, P., Xiang, H. J, Gong, X. G., Walsh, A., Wei, S. H., Intrinsic instability of the hybrid halid perovskite semiconductor $CH_3NH_3PbI_3$. *Chin. Phys. Lett.* **35**, 036104 (2018).
9. Syzgantseva, O. A., Saliba, M., Gratzel, M. & Rothlisberger, U. Stabilization of the Perovskite Phase of Formamidinium Lead Triiodide by Methylammonium, Cs, and/or Rb Doping. *J. Phys. Chem. Lett.* **8**, 1191-1196 (2017).
10. Rehman, W. *et al.* Photovoltaic mixed-cation lead mixed-halide perovskites: links between crystallinity, photo-stability and electronic properties. *Energ. Environ. Sci.* **10**, 361-369 (2017).
11. Li, L. *et al.* Precise composition tailoring of mixed-cation hybrid perovskites for efficient solar cells by mixture design methods. *ACS Nano* **11**, 8804-8813 (2017).
12. Kubicki, D. J. *et al.* Phase segregation in Cs-, Rb- and K-doped mixed-cation $(MA)_x(FA)_{1-x}PbI_3$ hybrid perovskites from solid state NMR. *J. Am. Chem. Soc.* **139**, 14173-14180 (2017).
13. Yi, C. *et al.* Entropic stabilization of mixed A-cation $ABX_3$ metal halide perovskites for high performance perovskite solar cells. *Energ. Environ. Sci.* **9**, 656-662 (2016).
14. Saliba, M. *et al.* Cesium-containing triple cation perovskite solar cells: improved stability, reproducibility and high efficiency. *Energ. Environ. Sci.* **9**, 1989-1997 (2016).
15. Saliba, M. *et al.* Incorporation of rubidium cations into perovskite solar cells improves photovoltaic performance. *Science* **354**, 206-209 (2016).
16. McMeekin, D. P. *et al.* A mixed-cation lead mixed-halide perovskiteabsorber for tandem solar cell. *Science* **351**, 5 (2016).
17. Li, Z. *et al.* Stabilizing perovskite structures by tuning tolerance factor: formation of formamidinium and cesium lead iodide solid-state alloys. *Chem. Mater.* **28**, 284-292 (2015).
18. Lee, J.-W. *et al.* Formamidinium and cesium hybridization for photo- and moisture-stable perovskite solar cell. *Adv. Energ. Mater.* **5**, 1501310 (2015).
19. Pellet, N. *et al.* Mixed-organic-cation perovskite photovoltaics for enhanced solar-light harvesting. *Angew. Chem. Int. Ed. Engl.* **53**, 3151-3157 (2014).
20. Volonakis, G., Haghighirad, A. A., Snaith, H. J. & Giustino, F. Route to Stable lead-free double perovskites with the electronic structure of $CH_3NH_3PbI_3$: a case for mixed-cation $[Cs/CH_3NH_3/CH(NH_2)_2]_2InBiBr_6$. *J. Phys. Chem. Lett.* **8**, 3917-3924 (2017).
21. The, D. *et al.* Rubidium multication perovskite with optimized bandgap for perovskite-silicon tandem with over 26% efficiency. *Adv. Energ. Mater.* **7** (2017).
22. Giustino, F. & Snaith, H. J. Toward lead-free perovskite solar cells. *ACS Energ. Lett.* **1**, 1233-1240 (2016).
23. McClure, E. T., Ball, M. R., Windl, W. & Woodward, P. M. $Cs_2AgBiX_6$(X = Br, Cl): new visible light absorbing, lead-free halide perovskite semiconductors. *Chem. Mater.* **28**, 1348-1354 (2016).
24. Slavney, A. H., Hu, T., Lindenberg, A. M. & Karunadasa, H. I. A Bismuth-halide double perovskite with long carrier recombination lifetime for photovoltaic applications. *J. Am. Chem. Soc.* **138**, 2138-2141 (2016).
25. Jain, A., Voznyy, O. & Sargent, E. H. High-throughput screening of lead-free perovskite-like materials for optoelectronic applications. *J. Phys. Chem. C.* **121**, 7183-7187 (2017).
26. Zhao, X. G. *et al.* Design of lead-free inorganic halide perovskites for solar cells via cation-transmutation. *J. Am. Chem. Soc.* **139**, 2630-2638 (2017).
27. M., G. V. *Naturwissenschaften* **14**, 477-485 (1926).

28. Travis, W., Glover, E. N. K., Bronstein, H., Scanlon, D. O. & Palgrave, R. G. On the application of the tolerance factor to inorganic and hybrid halide perovskites: a revised system. *Chem. Sci.* **7**, 4548-4556 (2016).
29. Sun, Q. D., Yin, W. J. Thermodynamic stability trend of cubic perovskites. *J. Am. Chem. Soc.*, **139**, 14905-14908 (2017).
30. Isayev, O. *et al.* Universal fragment descriptors for predicting properties of inorganic crystals. *Nature Commun.* **8**, 15679-15686 (2017).
31. Ghiringhelli, L. M., Vybiral, J., Levchenko, S. V., Draxl, C., Scheffler, M. Big data of materials science: critical role of the descriptor. *Phys. Rev. Lett.* **114**, 105503 (2015).
32. Fischer, C. C., Tibbetts, K. J., Morgan, D., Ceder, G. Predicting crystal structure by merging data mining with quantum mechanics. *Nature Mater.* **5**, 641-646 (2006).
33. Ahneman, D. T., Estrada, J. G., Lin, S., Dreher, S. D. & Doyle, A. G. Predicting reaction performance in C-N cross-coupling using machine learning. *Science (New York, N.Y.)* (2018).
34. Esteva, A. *et al.* Dermatologist-level classification of skin cancer with deep neural networks. *Nature* **542**, 115-118 (2017).
35. Silver, D. *et al.* Mastering the game of Go with deep neural networks and tree search. *Nature* **529**, 484-489 (2016).
36. Meyer, G. The synthesis and structures of comlex rare-earth halides. *Prog. Solid St. Chem.* **14**, 141-219 (1982).
37. Volonakis, G. *et al.* $Cs_2InAgCl_6$: A new lead-free halide double perovskite with direct band gap. *J. Phys. Chem. Lett.* **8**, 772-778 (2017).
38. Lamjed Debbichi *et al.* Mixed valence perovskite $Cs_2Au_2I_6$: a potential material for thin-film Pb-free photovoltaic cells with ultrahigh efficiency. *Adv. Mater.*, **2018**, 1707001 (2017).
39. Xiao, Z., Du, K.-Z., Meng, W., Mitzi, D. B., Yan, Y. Chemical origin of the stability difference between copper(I)- and silver(I)-based halide double perovskite. *Angew. Chem. Int. Ed. Engl.* **129**, 12275-12279 (2017).
40. Xiao, Z. *et al.* Intrinsic instability of $Cs_2In(I)M(III)X_6$ (M = Bi, Sb; X = Halogen) double perovskites: a combined density functional theory and experimental study. *J. Am. Chem. Soc.* **139**, 6054-6057 (2017).
41. Liao, W. *et al.* Fabrication of efficient low-bandgap perovskite solar cells by combining formamidinium tin iodide with methylammonium lead iodide. *J. Am. Chem. Soc.* **138**, 12360-12363 (2016).
42. Yang, Z. *et al.* Stable low-bandgap Pb-Sn binary perovskites for tandem solar cells. *Adv. Mater.* **28**, 8990-8997 (2016).
43. Eperon, G. E. *et al.* Perovskite-perovskite tandem photovoltaics with optimized band gaps. *Science* **354**, 861-865 (2016).
44. Gratzel, M. The rise of highly efficient and stable perovskite solar cells. *Acc. Chem. Res.* **50**, 487-491 (2017).
45. Ono, L. K., Juarez-Perez, E. J. & Qi, Y. Progress on perovskite materials and solar cells with mixed cations and halide anions. *Acs. Appl. Mater. Inter.* **9**, 30197-30246 (2017).
46. Perdew, J. P., Burke, K. & Ernzerhof, M. Generalized gradient approximation made simple. *Phys. Rev. Lett.* **77**, 3865-3868 (1996).
47. Kresse, G. & Joubert, D. From ultrasoft pseudopotentials to the projector augmented-wave method. *Phys Rev B* **59**, 1758-1775 (1999).
48. Murphy, K. P. Machine learning: a probabilistic perspective, The MIT Press, Chapter 14.4.3, 492-493 (2012)
49. Pedregosa, F. *et al.* Scikit-learn: machine learning in Python. *J. Mach. Learn. Res.* **12**, 2825-2830 (2011).

# Stability Engineering of Halide Perovskite via Machine Learning

Zhenzhu Li[1,2], Qichen Xu[1,2], Qingde Sun[1,2], Zhufeng Hou[3], Wan-Jian Yin[*,1,2]

[1]Soochow Institute for Energy and Materials InnovationS (SIEMIS), College of Physics, Optoelectronics and Energy & Collaborative Innovation Center of Suzhou Nano Science and Technology, Soochow University, Suzhou 215006, China

[2]Key Laboratory of Advanced Carbon Materials and Wearable Energy Technologies of Jiangsu Province, Soochow University, Suzhou 215006, China

[3]Center for Materials research by Information Integration, National Institute for Materials Science, 1-2-1 Sengen, Tsukuba, Ibaraki 305-0047, Japan

**Table S1.** Double halide perovskites in recent experimental attempt for synthesis. ✓ and ✗ mean stable and unstable respectively. Table lists the experimental, our DFT calculated, our ML-predicted and other DFT calculated results. The reference are provided. For other DFT calculated results, the decomposition pathways are also provided. (unit: meV/atom)

| **Double Perovskites** | **Exp. (✓ or ✗)** | **DFT cal. $\Delta H_D$ (✓ or ✗)** | **ML pre. $\Delta H_D$ (✓ or ✗)** | **DFT-$\Delta H_D$ Others (✓ or ✗)** |
|---|---|---|---|---|
| $Rb_2LiInCl_6$ | ✓[11] | | 56.45 (✓) | |
| $Cs_2LiInCl_6$ | ✓[11] | | 84.52 (✓) | |
| $In_2LiInCl_6$ | ✓[11] | | -235.53 (✗) | |
| $Rb_2NaInCl_6$ | ✓[11] | 95.49 (✓) | 54.60 (✓) | 26[7a] |
| $Cs_2NaInCl_6$ | ✓[12] | 111.10 (✓) | 93.59 (✓) | 30[7a] |
| $Cs_2NaInBr_6$ | ✓[11] | 25.83 (✓) | 51.90 (✓) | 9[7a] |
| $Cs_2NaSbCl_6$ | ✓[12] | 63.48 (✓) | 91.01 (✓) | 83[3a] |
| $Cs_2NaBiCl_6$ | ✓[13] | 94.58 (✓) | 87.06 (✓) | 114[3a] |
| $Cs_2NaTlCl_6$ | ✓[12] | | 95.44 (✓) | |
| $Cs_2NaFeCl_6$ | ✓[12] | | 63.24 (✓) | |
| $Cs_2NaTiCl_6$ | ✓[12] | | 81.64 (✓) | |
| $Cs_2KInCl_6$ | ✓[11] | 97.30 (✓) | 87.60 (✓) | 26[7a] |
| $Cs_2KBiCl_6$ | ✓[14] | 81.26 (✓) | 76.10 (✓) | 102[3a] |
| $Cs_2RbBiCl_6$ | ✓[15] | 68.39 (✓) | 79.46 (✓) | 90[3a] |
| $Cs_2AgInCl_6$ | ✓[16] | 133.39 (✓) | 94.78 (✓) | 116[1a] 89[4a] 19[4b] 47[4n] 8[4o] 7[4p] |
| $Cs_2AgSbCl_6$ | ✓[17] | 75.61 (✓) | 92.50 (✓) | 94[4a] 16[4b] 5[4o] 4[4p] 2[6x] 95[3a] |
| $Cs_2AgSbBr_6$ | ✓[18] | 56.04 (✓) | 51.39 (✓) | 83[3a] 79[4a] 8[4b] -2[4o] -3[4p] -10[6x] |
| $Cs_2AgBiCl_6$ | ✓[19] | 102.13 (✓) | 85.79 (✓) | 121[4a] 27 [4b] 16[4o] 15[4p] 12[6x] 121[3a] |
| $Cs_2AgBiBr_6$ | ✓[20] | 42.70 (✓) | 44.37 (✓) | 67[4a] 15[4b] 5[4e] 4[4p] -3[6x] 69[3a] |
| $Cs_2AgBiI_6$ | ✓[21] | -3.43 (✗) | 13.91 (✓) | 10[4a]-23[4b]-28[4e] -28[4g] -12[4n] -24[6g] 11[3a] |
| $Cs_2AgFeCl_6$ | ✓[22] | | 66.31 (✓) | |
| $Cs_2AgAuCl_6$ | ✓[23] | | 96.07 (✓) | |
| $K_2AuAuI_6$ | ✓[24] | | -61.77 (✗) | |
| $Rb_2AuAuBr_6$ | ✓[24] | | 0.62 (✓) | |
| $Rb_2AuAuI_6$ | ✓[24] | | -18.16 (✗) | |
| $Cs_2AuAuCl_6$ | ✓[24] | | 89.02 (✓) | |
| $Cs_2AuAuBr_6$ | ✓[24] | | 43.38 (✓) | |
| $Cs_2AuAuI_6$ | ✓[24] | | 21.89 (✓) | |
| $Cs_2InInCl_6$ | ✓[11] | 116.39 (✓) | 93.77 (✓) | |
| $Cs_2TlBiCl_6$ | ✓[25] | 84.17 (✓) | 74.97 (✓) | 112[3a] |
| $Cs_2TlTlCl_6$ | ✓[26] | | 90.42 (✓) | |
| $Cs_2CuInCl_6$ | ✗[4] | 111.04 (✓) | 84.57 (✓) | 77[1a] 50[4a] -20[4b] -41[4g] 10[4h] -14[4i] |
| $Cs_2CuInBr_6$ | ✗[4] | 34.77 (✓) | 48.64 (✓) | 23[1a] 22[4a] -33[4b] -50[4g] -9[4h] -28[4i] |
| $Cs_2AgInBr_6$ | ✗[4,8] | 59.85 (✓) | 54.33 (✓) | 56[1a] 56[4a] 0[4b] 17[4n] -10[4o] -11[4p] |
| $Cs_2AgInI_6$ | ✗[4] | 14.85 (✓) | 30.95 (✓) | 3[1a] 3[4a] -43[4j] -19[4n] -54[4p] |
| $Cs_2InBiCl_6$ | ✗[2] | 74.38 (✓) | 83.90 (✓) | 110[2a] 15[2b] 85[2c] 54[2d] 15[2e] -17[2f] -24[5f] 110[3a] |
| $Cs_2InBiBr_6$ | ✗[2,7] | 49.14 (✓) | 42.07 (✓) | 64[2a] 12[2b] 71[2c] 54[2d] 16[2e] -2[2f] -1[5f] 65[3a] |
| $Cs_2InBiI_6$ | ✗[2] | 20.48 (✓) | 13.59 (✓) | 32[2a] -1[2b] 60[2c] 50[2d] 15[2e] 4[2f] -12[5b] 32[3a] |
| $Cs_2TlBiBr_6$ | ✗[9,10] | 34.97 (✓) | 22.74 (✓) | 51[3a] |

*For reference in the table, numbers indicate reference numbers and alphabets indicate decomposition pathways. For example, ref. [2a] means the value was based on decomposition pathway [a] cited from ref. [2]. The decomposition pathways and references are appended below.

**Decomposition Pathways:**

[a] A2M(I)M(III)X6→2AX+ M(I)X+ M(III)X3
[b] A2M(I)M(III)X6→M(I)X +1/2A3 M(III)2X9+1/2 AX
[c] A2M(I)M(III)X6→1/2 AX + M(I)X3+1/3 M(III)X3+2/3 M(III)
[d] A2M(I)M(III)X6→3/2 AX+ M(I)X3+1/6A3 M(III)2X9+2/3 M(III)
[e] A2M(I)M(III)X6→1/2A3 M(I)2X9+1/2 AX +1/3 M(III)X3+2/3 M(III)
[f] A2M(I)M(III)X6→1/2A3 M(I)2X9+1/6A3 M(III)2X9+2/3 M(III)
[g] A2M(I)M(III)X6→1/2 A M(I)2X3 + 1/2 A3 M(III)2X9
[h] A2M(I)M(III)X6→1/2 A3M(I)2X5 +1/2AX + M(III)X3
[i] A2M(I)M(III)X6→1/2 A3M(I)2X5 + 1/6 A3 M(III)2X9 + 2/3 M(III)X3
[j] A2M(I)M(III)X6→AX + M(I)X + AM(III)X4
[k] A2M(I)M(III)X6→1/2 AX + 1/2 AM(I)2X3 + AM(III)X4
[l] A2M(I)M(III)X6→1/3 A3M(I)2X5 + 1/3 M(I)X + AM(III)X4
[m] A2M(I)M(III)X6→1/4 A M(I)2X3 +1/4 A3M(I)2X5 + AM(III)X4
[n] A2M(I)M(III)X6→A2M(I)2X3 + M(III)X3
[o] A2M(I)M(III)X6→1/4 A M(I)2X3 + 3/4 M(I)X + 1/2 A3 M(III)2X9
[p] A2M(I)M(III)X6→1/2 AM(I)2X3 + 1/2 M(I)X + 1/2 A3M(III)2X9
[q] A2M(I)M(III)X6→1/3A2M(I)X3 + 1/3 AM(I)2X3 + AM(III)X4
[r] A2M(I)M(III)X6→1/8A4M(I)5X9 + 3/8 M(I)X + 1/2 A3M(III)2X9
[s] A2M(I)M(III)X6→7/15M(III)X3+1/15M(I)5X9+2/3A3M(I)X6+8/15M(III)
[t] A2M(I)M(III)X6→3/22M(III)6X7+2/11M(III)X3+1/6M(I)2X3+2/3A3M(I)X6
[u] A2M(I)M(III)X6→1/3M(III)X3+A2M(I)X5+2/3M(III)
[v] A2M(I)M(III)X6→2/3A3M(III)X6+ M(I)X2+1/3 M(III)
[w] A2M(I)M(III)X6→1/6 A3M(III)2X9+ 1/2AX + AM(I)X4 +2/3 M(III)
[x] A2M(I)M(III)X6→1/2 M(I)+ 1/2 A3M(III)2X9+1/2 AM(I)X3

[1] Zhao, X.-G.; Yang, D.; Sun, Y.; Li, T.; Zhang, L.; Yu, L.; Zunger, A. J. Am. Chem. Soc. 2017, 139 (19), 6718-6725.
[2] Xiao, Z.; Du, K.-Z.; Meng, W.; Wang, J.; Mitzi, D. B.; Yan, Y. J. Am. Chem. Soc. 2017, 139 (17), 6054-6057.
[3] Zhao, X.-G.; Yang, J.-H.; Fu, Y.; Yang, D.; Xu, Q.; Yu, L.; Wei, S.- H.; Zhang, L. J. Am. Chem. Soc. 2017, 139, 2630−2638.
[4] Xiao, Z.; Du, K.-Z.; Meng, W.; Mitzi, D. B.; Yan, Y. Angewandte Chemie-International Edition 2017, 56 (40), 12107-12111.
[5] Volonakis, G.; Haghighirad, A. A.; Snaith, H. J.; Giustino, F. The Journal of Physical Chemistry Letters 2017, 8 (16), 3917-3924.
[6] Filip, M. R.; Liu, X.; Miglio, A.; Hautier, G.; Giustino, F. The Journal of Physical Chemistry C 2017, 122 (1), 158-170.
[7] Jain, A.; Voznyy, O.; Sargent, E. H. The Journal of Physical Chemistry C 2017, 121 (13), 7183-7187.
[8] Xu, J.; Liu, J.-B.; Liu, B.-X.; Huang, B. The Journal of Physical Chemistry Letters 2017, 8 (18), 4391-4396.
[9] Xiao, Z.; Yan, Y.; Hosono, H.; Kamiya, T. The Journal of Physical Chemistry Letters 2018, 9 (1), 258-262.
[10] Savory, C. N.; Walsh, A.; Scanlon, D. O. ACS energy letters 2016, 1 (5), 949-955.
[11] Knop, O.; Cameron, T. S.; Adhikesavalu, D.; Vincent, B. R.; Jenkins, J. A. Can. J. Chem. 1987, 65 (7), 1527-1556.
[12] Morss, L. R.; Siegal, M.; Stenger, L.; Edelstein, N. Inorg. Chem. 1970, 9, 1771-1775.
[13] Meyer, G.; Hwu, S. J.; Corbett, J. D. Z. Anorg. Allg. Chem. 1986, 535, 208-212.
[14] Barbier, P.; Drache, M.; Mairesse, G.; Ravez, J. J. Solid State Chem. 1982, 42, 130-135.
[15] Tomaszewski, P. E. Phase Transitions 1992, 38 (3), 127.
[16] Volonakis G, Haghighirad AA, Milot RL, Sio WH, Filip MR, Wenger B, et al. Cs2InAgCl6: A new lead-free halide double perovskite with direct band gap. J Phys Chem Lett 2017, 8(4): 772-778.
[17] Tran, T. T.; Panella, J. R.; Chamorro, J. R.; Morey, J. R.; McQueen, T. M. Mater. Horiz. 2017, 4, 688-693.
[18] Du, K.; Meng, W.; Wang, X.; Yan, Y.; Mitzi, D. B. Angew. Chem. Int. Ed. 2017.
[19] Deng Z, Wei F, Sun S, Kieslich G, Cheetham AK, Bristowe PD. Exploring the properties of lead-free hybrid double perovskites using a combined computational-experimental approach. J Mater Chem A 2016, 4(31): 12025-12029.
[20] Slavney AH, Hu T, Lindenberg AM, Karunadasa HI. A bismuth-halide double perovskite with long carrier recombination lifetime for photovoltaic applications. J Am Chem Soc 2016, 138(7): 2138-2141.
[21] Creutz, S. E.; Crites, E. N.; Michael, C.; Gamelin, D. R. 2018.
[22] Flerov, I.; Gorev, M.; Aleksandrov, K.; Tressaud, A.; Grannec, J.; Couzi, M. Materials Science and Engineering: R: Reports 1998, 24 (3), 81-151.
[23] Elliott, N.; Pauling, L. J. Am. Chem. Soc. 1938, 60, 1846.
[24] Kojima, N. Bull. Chem. Soc. Jpn. 2000, 73, 1445.
[25] Beznosikov, B. V.; Mosyul, S. V. Sov. Phys. Crystallogr. 1978, 23, 622.
[26] Retuerto, M.; Emge, T.; Hadermann, J.; Stephens, P. W.; Li, M.R.; Yin, Z. P.; Croft, M.; Ignatov, A.; Zhang, S. J.; Yuan, Z.; et al. Chem.Mater. 2013, 25, 4071.

**Table S2.** Experimental compositions of mixed halide perovskites marked in Figure 5.

| | | ① | ② | ③ | ④ | ⑤ | ⑥ | ⑦ | ⑧ | ⑨ | ⑩ | ⑪ | ⑫ | ⑬ | ⑭ | ⑮ | ⑯ | ⑰ | ⑱ |
|---|---|---|---|---|---|---|---|---|---|---|---|---|---|---|---|---|---|---|---|
| **A** | **Rb** | | | | | | | | | 0.05 | | | | | | | | | |
| | **Cs** | | | | | 0.10 | 0.10 | 0.15 | 0.20 | | | | | | | | | | |
| | **MA** | 0.13 | 0.25 | 0.60 | 0.70 | 0.90 | | | | | 1.00 | 1.00 | 1.00 | 1.00 | 1.00 | 1.00 | 1.00 | 1.00 | 1.00 |
| | **FA** | 0.87 | 0.75 | 0.40 | 0.30 | | 0.90 | 0.85 | 0.80 | 0.95 | | | | | | | | | |
| **B** | **Sn** | | | | | | | | | | 0.25 | 0.50 | 0.75 | 0.10(Hg) | | | | | |
| | **Pb** | 1.00 | 1.00 | 1.00 | 1.00 | 1.00 | 1.00 | 1.00 | 1.00 | 1.00 | 0.75 | 0.50 | 0.25 | 0.90 | 1.00 | 1.00 | 1.00 | 1.00 | 1.00 |
| **X** | **Cl** | | | | | | | | | | | | | | | | | | |
| | **Br** | | | | | | | | | | | | | | 1.80 | 0.60 | 2.10 | 2.00 | 2.50 |
| | **I** | 3.00 | 3.00 | 3.00 | 3.00 | 3.00 | 3.00 | 3.00 | 3.00 | 3.00 | 3.00 | 3.00 | 3.00 | 3.00 | 1.20 | 2.40 | 0.90 | 1.00 | 0.50 |
| **Ref** | | [1] | [2] | [3] | [4] | [5] | [7] | [8] | [9] | [10] | [38] | [37] | [38] | [39] | [11,12] | [12] | [12] | [13] | [13] |
| | | ⑲ | ⑳ | ㉑ | ㉒ | ㉓ | ㉔ | ㉕ | ㉖ | ㉗ | ㉘ | ㉙ | ㉚ | ㉛ | ㉜ | ㉝ | ㉞ | ㉟ | ㊱ |
| **A** | **Rb** | | | | | | | | | | | | | | | | | 0.05 | 0.05 |
| | **Cs** | | | 1.00 | | | | 0.25 | 0.20 | 0.17 | | | | | 0.17 | 0.10 | 0.10 | 0.10 | 0.10 |
| | **MA** | 1.00 | 1.00 | | | 0.40 | 0.50 | | | | 0.16 | 0.20 | 0.15 | 0.33 | | 0.90 | 0.10 | | 0.14 |
| | **FA** | | | | 1.00 | 0.60 | 0.50 | 0.75 | 0.80 | 0.83 | 0.84 | 0.80 | 0.85 | 0.67 | 0.83 | | 0.80 | 0.85 | 0.71 |
| **B** | **Sn** | | | | | 0.60 | 0.25 | 0.50 | | | | | | | | | | | |
| | **Pb** | 1.00 | 1.00 | 1.00 | 1.00 | 0.40 | 0.75 | 0.50 | 1.00 | 1.00 | 1.00 | 1.00 | 1.00 | 1.00 | 1.00 | 1.00 | 1.00 | 1.00 | 1.00 |
| **X** | **Cl** | 0.45 | 0.75 | | | | | | | | | | | | | | | | |
| | **Br** | 2.55 | 2.25 | 1.00 | 2.00 | | | | 0.16 | 1.20 | 0.48 | 0.60 | 0.45 | 0.50 | 1.50 | 0.30 | 0.51 | 0.51 | 1.00 |
| | **I** | | | 2.00 | 1.00 | 3.00 | 3.00 | 3.00 | 2.84 | 1.80 | 2.52 | 2.40 | 2.55 | 2.50 | 1.50 | 2.70 | 2.49 | 2.49 | 2.00 |
| **Ref** | | [19] | [19] | [20] | [17] | [41] | [42] | [43] | [22] | [24] | [23] | [25] | [26] | [27] | [43] | [28] | [32] | [34] | [36] |

[1] Binek, A.; Hanusch, F. C.; Docampo, P.; Bein, T. Stabilization of the Trigonal HighTemperature Phase of Formamidinium Lead Iodide. J. Phys. Chem. Lett. 2015, 6, 1249-1253.
[2] Ji, F.; Wang, L.; Pang, S.; Gao, P.; Xu, H.; Xie, G.; Zhang, J.; Cui, G. A Balanced Cation Exchange Reaction Toward Highly Uniform and Pure Phase FA1-xMAxPbI3 Perovskite Films. J. Mater. Chem. A 2016, 4, 14437-14443.
[3] Deng, Y.; Dong, Q.; Bi, C.; Yuan, Y.; Huang, J. Air-Stable, Efficient Mixed-Cation Perovskite Solar Cells with Cu Electrode by Scalable Fabrication of Active Layer. Adv. Energy Mater. 2016, 6, 1600372.
[4] Li, G.; Zhang, T.; Guo, N.; Xu, F.; Qian, X.; Zhao, Y. Ion-Exchange-Induced 2D–3D Conversion of HMA1−xFAxPbI3Cl Perovskite into a High-Quality MA1−xFAxPbI3 Perovskite.Angew. Chem. Int. Ed. 2016, 55, 13460-13464.
[5] Choi, H.; Jeong, J.; Kim, H.-B.; Kim, S.; Walker, B.; Kim, G.-H.; Kim, J. Y. CesiumDoped Methylammonium Lead Iodide Perovskite Light Absorber for Hybrid Solar Cells. Nano Energy 2014, 7, 80-85.
[6] Niu, G.; Li, W.; Li, J.; Liang, X.; Wang, L. Enhancement of Thermal Stability for Perovskite Solar Cells through Cesium Doping. RSC Adv. 2017, 7, 17473-17479.
[7] Lee, J.-W.; Kim, D.-H.; Kim, H.-S.; Seo, S.-W.; Cho, S. M.; Park, N.-G. Formamidinium and Cesium Hybridization for Photo- and Moisture-Stable Perovskite Solar Cell. Adv. Energy Mater. 2015, 5, 1501310.
[8] Li, Z.; Yang, M.; Park, J.-S.; Wei, S.-H.; Berry, J. J.; Zhu, K. Stabilizing Perovskite Structures by Tuning Tolerance Factor: Formation of Formamidinium and Cesium Lead Iodide Solid-State Alloys. Chem. Mater. 2016, 28, 284-292.
[9] Yi, C.; Luo, J.; Meloni, S.; Boziki, A.; Ashari-Astani, N.; Gratzel, C.; Zakeeruddin, S. M.; Rothlisberger, U.; Gratzel, M. Entropic Stabilization of Mixed A-Cation ABX3 Metal Halide Perovskites for High Performance Perovskite Solar Cells. Energy Environ. Sci. 2016, 9, 656-662.
[10] Park, Y. H.; Jeong, I.; Bae, S.; Son, H. J.; Lee, P.; Lee, J.; Lee, C.-H.; Ko, M. J. Inorganic Rubidium Cation as an Enhancer for Photovoltaic Performance and Moisture Stability of HC(NH2)2PbI3 Perovskite Solar Cells. Adv. Funct. Mater. 2017, 27, 1605988.
[11] Sadhanala, A.; Deschler, F.; Thomas, T. H.; Dutton, S. E.; Goedel, K. C.; Hanusch, F. C.; Lai, M. L.; Steiner, U.; Bein, T.; Docampo, P. et al. Preparation of Single-Phase Films of CH3NH3Pb(I1–xBrx)3 with Sharp Optical Band Edges. J. Phys. Chem. Lett. 2014, 5, 2501-2505.
[12] Bischak, C. G.; Hetherington, C. L.; Wu, H.; Aloni, S.; Ogletree, D. F.; Limmer, D. T.; Ginsberg, N. S. Origin of Reversible Photoinduced Phase Separation in Hybrid Perovskites. Nano Lett. 2017, 17, 1028-1033.
[13] Brivio, F.; Caetano, C.; Walsh, A. Thermodynamic Origin of Photoinstability in the CH3NH3Pb(I1–xBrx)3 Hybrid Halide Perovskite Alloy. J. Phys. Chem. Lett. 2016, 7, 1083-1087.
[14] Lv, S.; Pang, S.; Zhou, Y.; Padture, N. P.; Hu, H.; Wang, L.; Zhou, X.; Zhu, H.; Zhang, L.; Huang, C. et al. One-Step, Solution-Processed Formamidinium Lead Trihalide (FAPbI(3-x)Clx) for Mesoscopic Perovskite-Polymer Solar Cells. Phys. Chem. Chem. Phys. 2014, 16, 19206- 19211.
[15] Wang, Z.; Zhou, Y.; Pang, S.; Xiao, Z.; Zhang, J.; Chai, W.; Xu, H.; Liu, Z.; Padture, N. P.; Cui, G. Additive-Modulated Evolution of HC(NH2)2PbI3 Black Polymorph for Mesoscopic Perovskite Solar Cells. Chem. Mater. 2015, 27, 7149-7155.
[16] Zhang, H.; Liao, Q.; Wang, X.; Hu, K.; Yao, J.; Fu, H. Controlled Substitution of Chlorine for Iodine in Single-Crystal Nanofibers of Mixed Perovskite MAPbI3–xClx. Small 2016, 12, 3780-3787.
[17] Rehman, W.; Milot, R. L.; Eperon, G. E.; Wehrenfennig, C.; Boland, J. L.; Snaith, H. J.; Johnston, M. B.; Herz, L. M. Charge-Carrier Dynamics and Mobilities in Formamidinium Lead Mixed-Halide Perovskites. Adv. Mater. 2015, 27, 7938-7944.
[18] Eperon, G. E.; Stranks, S. D.; Menelaou, C.; Johnston, M. B.; Herz, L. M.; Snaith, H. J. Formamidinium Lead Trihalide: A Broadly Tunable Perovskite for Efficient Planar Heterojunction Solar Cells. Energy Environ. Sci. 2014, 7, 982-988.
[19] Zhang, T.; Yang, M.; Benson, E. E.; Li, Z.; van de Lagemaat, J.; Luther, J. M.; Yan, Y.; Zhu, K.; Zhao, Y. A Facile Solvothermal Growth of Single Crystal Mixed Halide Perovskite CH3NH3Pb(Br1-xClx)3. Chem. Commun. 2015, 51, 7820-7823.
[20] Sutton, R. J.; Eperon, G. E.; Miranda, L.; Parrott, E. S.; Kamino, B. A.; Patel, J. B.; Hörantner, M. T.; Johnston, M. B.; Haghighirad, A. A.; Moore, D. T. et al. Bandgap-Tunable Cesium Lead Halide Perovskites with High Thermal Stability for Efficient Solar Cells. Adv. Energy Mater. 2016, 6, 1502458.
[21] Beal, R. E.; Slotcavage, D. J.; Leijtens, T.; Bowring, A. R.; Belisle, R. A.; Nguyen, W. H.; Burkhard, G. F.; Hoke, E. T.; McGehee, M. D. Cesium Lead Halide Perovskites with Improved Stability for Tandem Solar Cells. J. Phys. Chem. Lett. 2016, 7, 746-751.
[22] Yi, C.; Luo, J.; Meloni, S.; Boziki, A.; Ashari-Astani, N.; Gratzel, C.; Zakeeruddin, S. M.; Rothlisberger, U.; Gratzel, M. Entropic Stabilization of Mixed A-Cation ABX3 Metal Halide

Perovskites for High Performance Perovskite Solar Cells. Energy Environ. Sci. 2016, 9, 656-662.
[23] Chen, B.-X.; Li, W.-G.; Rao, H.-S.; Xu, Y.-F.; Kuang, D.-B.; Su, C.-Y. Large-Grained Perovskite Films via FAxMA1−xPb(IxBr1−x)3 Single Crystal Precursor for Efficient Solar Cells.Nano Energy 2017, 34, 264-270.
[24] Wang, Z.; McMeekin, D. P.; Sakai, N.; van Reenen, S.; Wojciechowski, K.; Patel, J. B.; Johnston, M. B.; Snaith, H. J. Efficient and Air-Stable Mixed-Cation Lead Mixed-Halide Perovskite Solar Cells with n-Doped Organic Electron Extraction Layers. Adv. Mater. 2017, 29, 1604186.
[25] Chen, L. C.; Tseng, Z. L.; Huang, J. K. A Study of Inverted-Type Perovskite Solar Cells with Various Composition Ratios of (FAPbI3)(1-x)(MAPbBr3)x. Nanomater. 2016, 6, 183.
[26] Sveinbjornsson, K.; Aitola, K.; Zhang, J.; Johansson, M. B.; Zhang, X.; Correa-Baena, J.- P.; Hagfeldt, A.; Boschloo, G.; Johansson, E. M. J. Ambient Air-Processed Mixed-Ion Perovskites for High-Efficiency Solar Cells. J. Mater. Chem. A 2016, 4, 16536-16545.
[27] Jacobsson, T. J.; Correa-Baena, J.-P.; Pazoki, M.; Saliba, M.; Schenk, K.; Gratzel, M.; Hagfeldt, A. Exploration of the Compositional Space for Mixed Lead Halogen Perovskites for High Efficiency Solar Cells. Energy Environ. Sci. 2016, 9, 1706-1724.
[28] Niu, G.; Yu, H.; Li, J.; Wang, D.; Wang, L. Controlled Orientation of Perovskite Films Through Mixed Cations Toward High Performance Perovskite Solar Cells. Nano Energy 2016,27, 87-94.
[29] Isikgor, F. H.; Li, B.; Zhu, H.; Xu, Q.; Ouyang, J. High Performance Planar Perovskite Solar Cells With A Perovskite Of Mixed Organic Cations And Mixed Halides, MA1-xFAxPbI3-yCly. J. Mater. Chem. A 2016, 4, 12543-12553.
[30] Chiang, C.-H.; Lin, J.-W.; Wu, C.-G. One-Step Fabrication of a Mixed-Halide Perovskite Film for a High-Efficiency Inverted Solar Cell and Module. J. Mater. Chem. A 2016, 4, 13525-13533
[31] Saliba, M.; Matsui, T.; Seo, J.-Y.; Domanski, K.; Correa-Baena, J.-P.; Nazeeruddin, M. K.; Zakeeruddin, S. M.; Tress, W.; Abate, A.; Hagfeldt, A. et al. Cesium-Containing Triple Cation Perovskite Solar Cells: Improved Stability, Reproducibility and High Efficiency. Energy
Environ. Sci. 2016, 9, 1989-1997.
[32] Matsui, T.; Seo, J.-Y.; Saliba, M.; Zakeeruddin, S. M.; Grätzel, M. Room-Temperature Formation of Highly Crystalline Multication Perovskites for Efficient, Low-Cost Solar Cells.Adv. Mater. 2017, 29, 1606258.
[33] Zhang, M.; Yun, J. S.; Ma, Q.; Zheng, J.; Lau, C. F. J.; Deng, X.; Kim, J.; Kim, D.; Seidel, J.; Green, M. A. et al. High-Efficiency Rubidium-Incorporated Perovskite Solar Cells by Gas Quenching. ACS Energy Lett. 2017, 2, 438-444.
[34] Saliba, M.; Matsui, T.; Domanski, K.; Seo, J. Y.; Ummadisingu, A.; Zakeeruddin, S. M.; Correa-Baena, J. P.; Tress, W. R.; Abate, A.; Hagfeldt, A. et al. Incorporation of Rubidium Cations Into Perovskite Solar Cells Improves Photovoltaic Performance. Science 2016, 354, 206- 209.
[35] Philippe, B.; Saliba, M.; Correa-Baena, J.-P.; Cappel, U. B.; Turren-Cruz, S.-H.; Grätzel, M.; Hagfeldt, A.; Rensmo, H. Chemical Distribution of Multiple Cation (Rb+, Cs+, MA+, and FA+) Perovskite Materials by Photoelectron Spectroscopy. Chem. Mater. 2017, 29, 3589-3596.
[36] Duong, T.; Wu, Y.; Shen, H.; Peng, J.; Fu, X.; Jacobs, D.; Wang, E.-C.; Kho, T. C.; Fong, K. C.; Stocks, M. et al. Rubidium Multication Perovskite with Optimized Bandgap for Perovskite-Silicon Tandem with over 26% Efficiency. Adv. Energy Mater. 2017, 7, 1700228.
[37] Y. Ogomi, A. Morita, S. Tsukamoto, T. Saitho, N. Fujikawa,Q. Shen, T. Toyoda, K. Yoshino, S. S. Pandey, T. Ma, J. Phys. Chem.Lett. 2014, 5, 1004.
[38] F. Hao, C. C. Stoumpos, R. P. H. Chang, M. G. Kanatzidis, J. Am.Chem. Soc. 2014, 136, 8094.
[39] L. A. Frolova, D. V. Anokhin, K. L. Gerasimov, N. N. Dremova,P. A. Troshin, J. Phys. Chem. Lett. 2016, 7, 4353.
[40] M. Zhang, H. Yu, M. Lyu, Q. Wang, J.-H. Yun, L. Wang, Chem.Commun. 2014, 50, 11727.
[41] W. Liao, D. Zhao, Y. Yu, N. Shrestha, K. Ghimire,C. R. Grice, C. Wang, Y. Xiao, A. J. Cimaroli,R. J. Ellingson, N. J. Podraza, K. Zhu, R.-G. Xiong and Y. Yan, J. Am. Chem. Soc., 2016, 138, 12360–12363.
[42] Z. Yang, A. Rajagopal, C.-C. Chueh, S. B. Jo, B. Liu, T. Zhao and A. K. Y. Jen, Adv. Mater., 2016, 28, 8990–8997.
[43] G. E. Eperon, T. Leijtens, K. A. Bush, R. Prasanna,M. D. McGehee and H. J. Snaith, et. al., Science, 2016,354, 861–865.

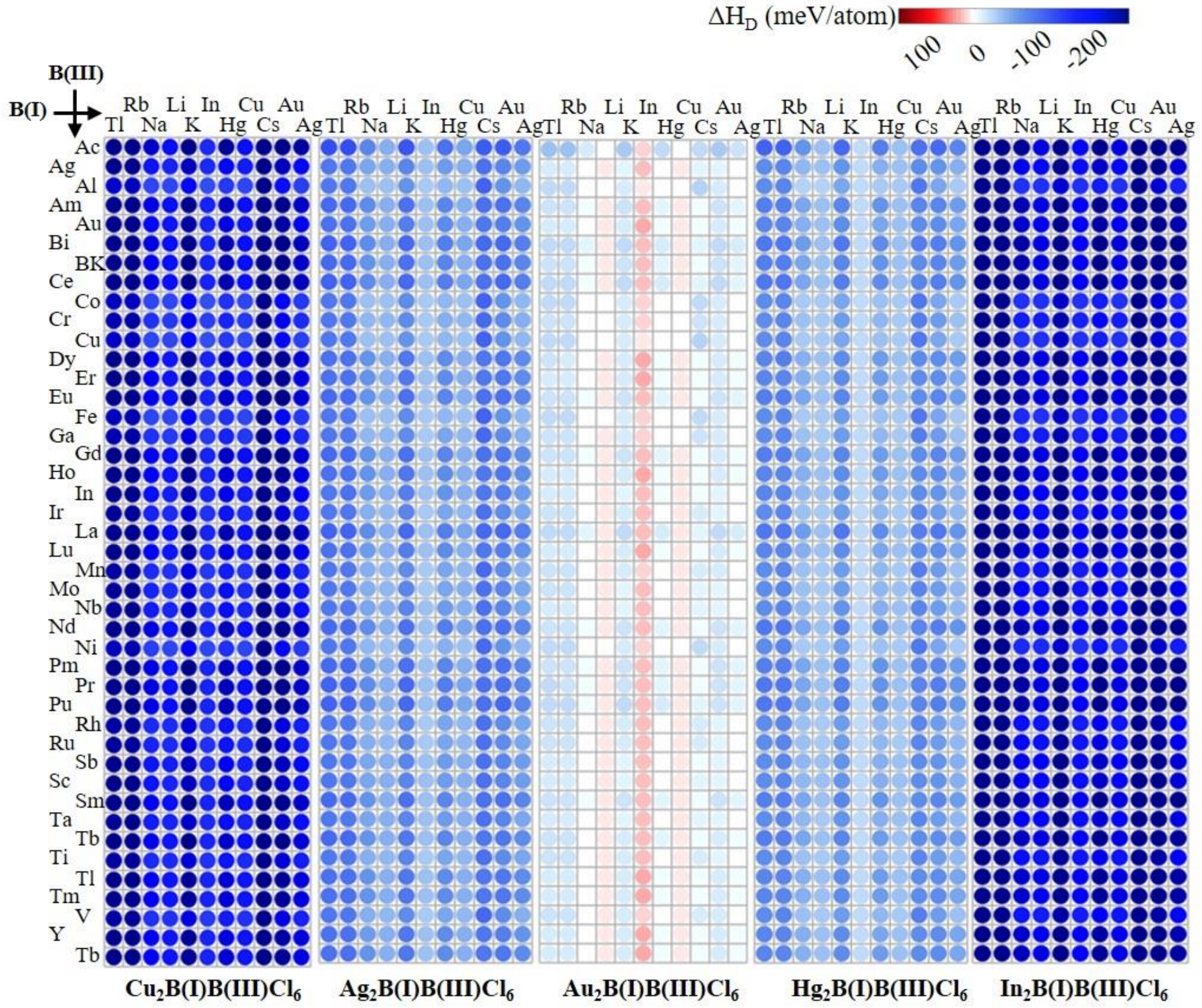

**Figure S1. Heat map of decomposition energies of 2365 perovskites with A=Cu, Ag, Au, Hg, In and X=Cl.** The red/blue color indicates the positive/negative decomposition energies and the absolute value is presented by color shades.

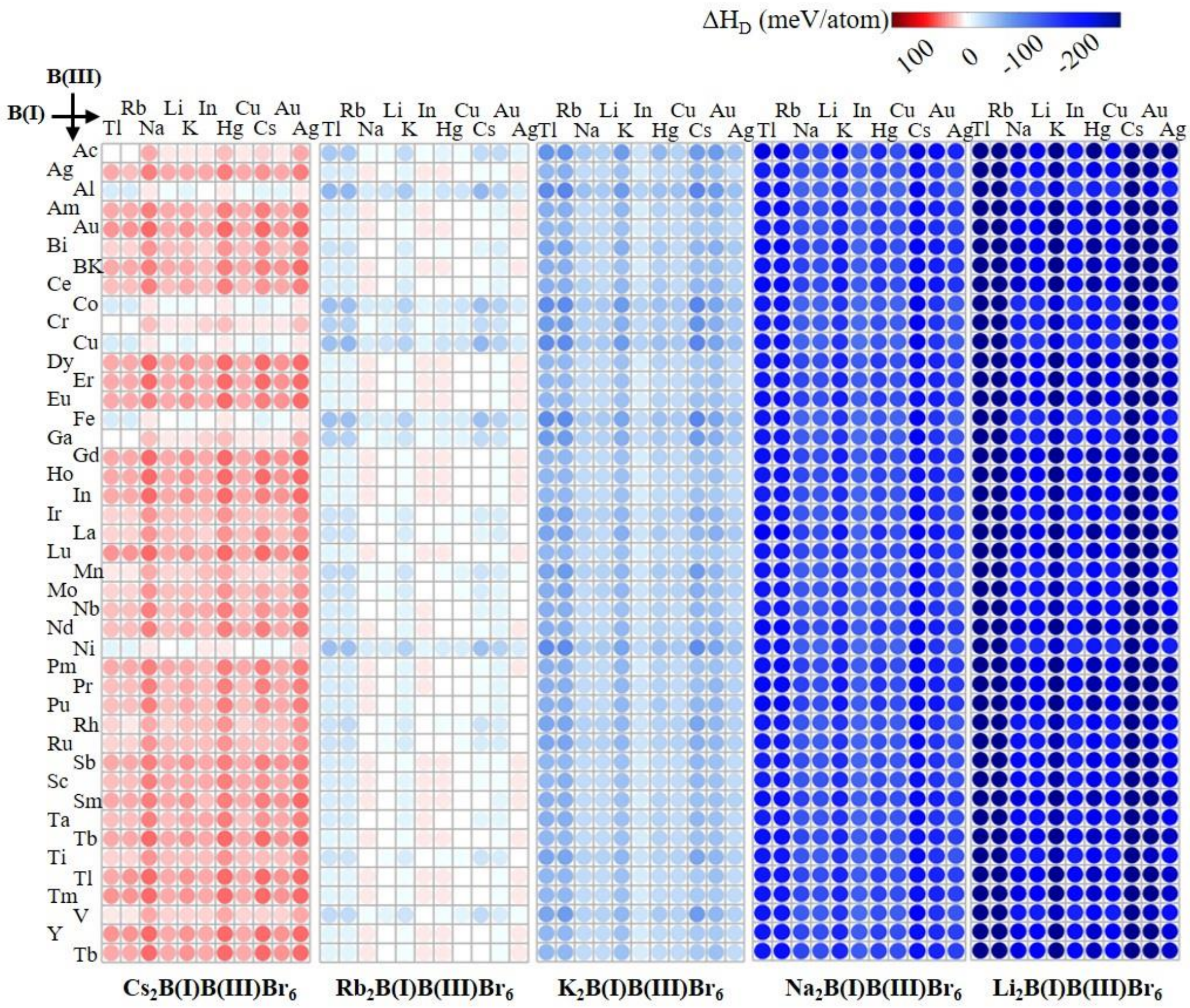

**Figure S2. Heat map of decomposition energies of 2365 perovskites with A=Cs, Rb, K, Na, Li and X=Br.** The red/blue color indicates the positive/negative decomposition energies and the absolute value is presented by color shades.

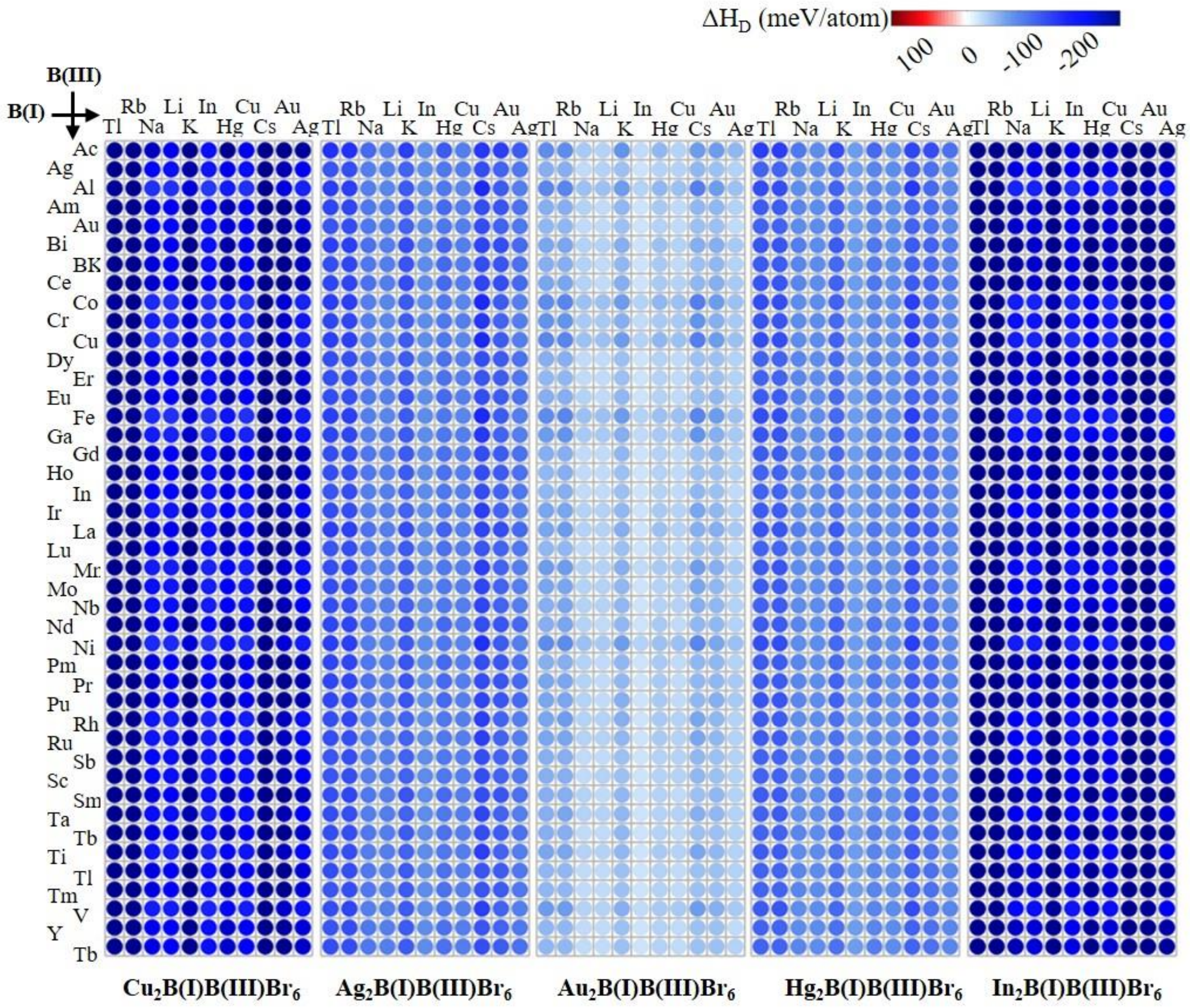

**Figure S3. Heat map of decomposition energies of 2365 perovskites with A=Cu, Ag, Au, Hg, In and X=Br.** The red/blue color indicates the positive/negative decomposition energies and the absolute value is presented by color shades.

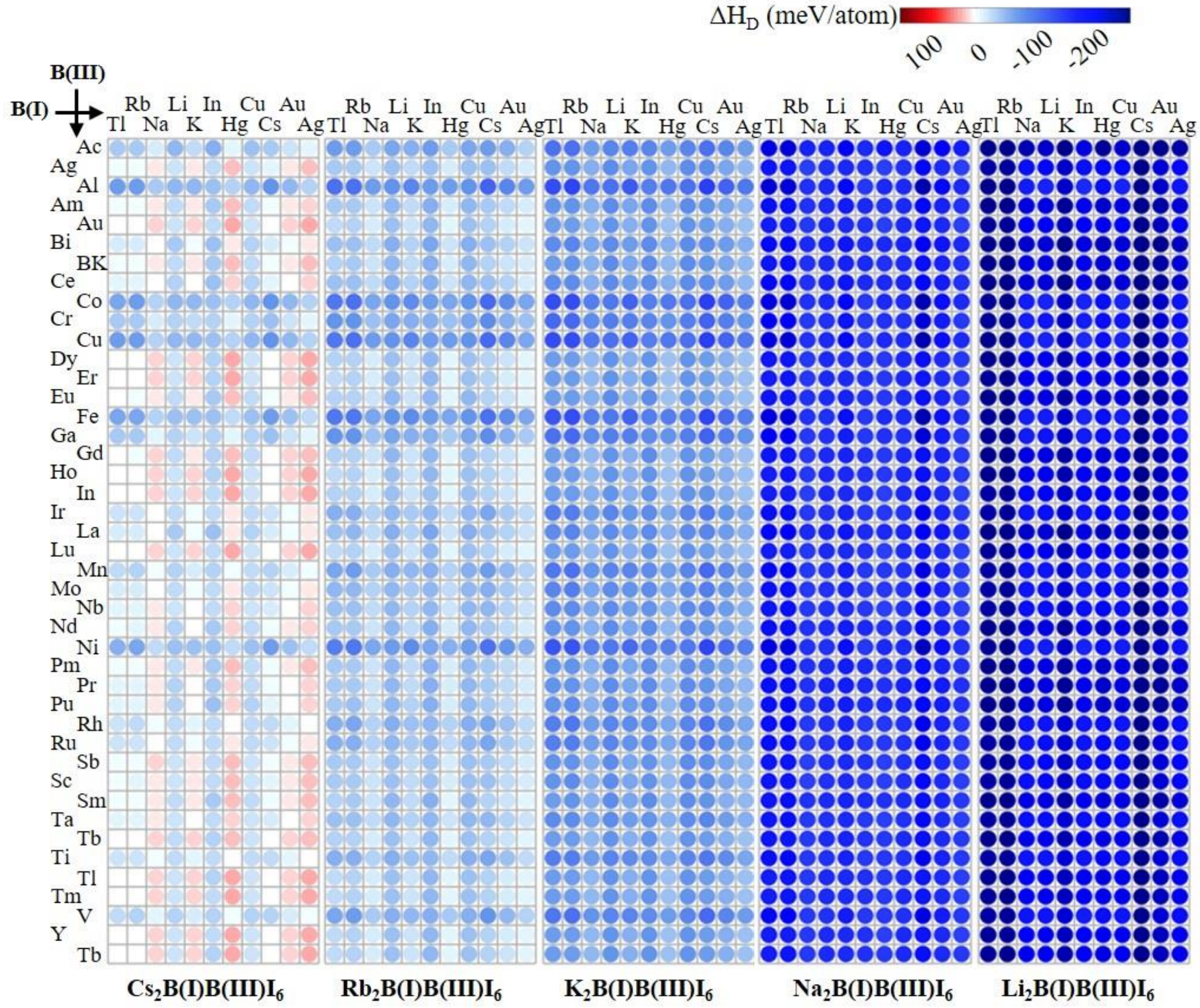

**Figure S4. Heat map of decomposition energies of 2365 perovskites with A=Cs, Rb, K, Na, Li and X=I.** The red/blue color indicates the positive/negative decomposition energies and the absolute value is presented by color shades.

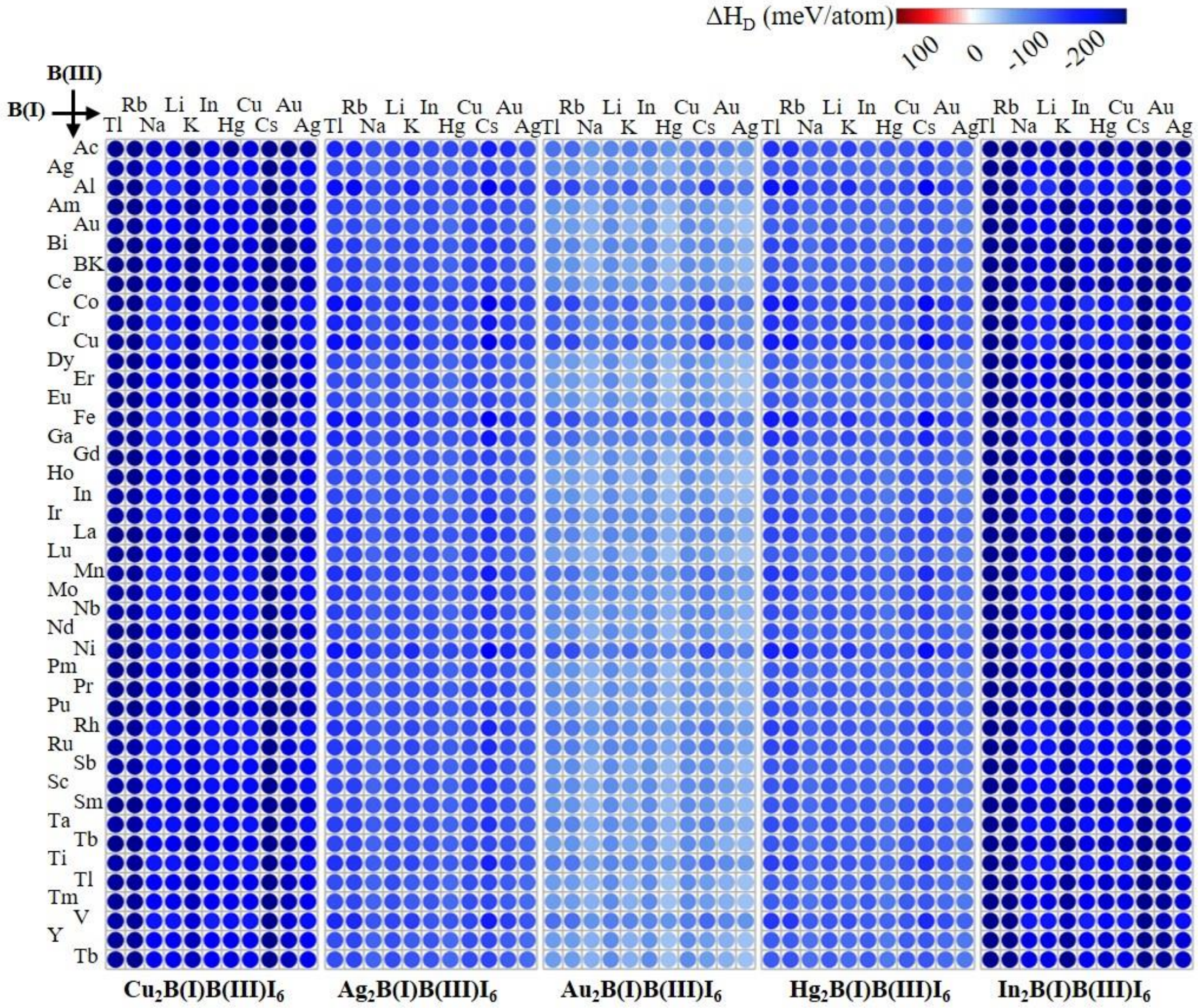

**Figure S5. Heat map of decomposition energies of 2365 perovskites with A=Cu, Ag, Au, Hg, In and X=Br.** The red/blue color indicates the positive/negative decomposition energies and the absolute value is presented by color shades.